\documentclass[conference]{IEEEtran}
\IEEEoverridecommandlockouts

\usepackage{cite}
\usepackage{amsmath,amssymb,amsfonts}
\usepackage{algorithmicx}
\usepackage{graphicx}
\usepackage{textcomp}
\usepackage{xcolor}
\def\BibTeX{{\rm B\kern-.05em{\sc i\kern-.025em b}\kern-.08em
    T\kern-.1667em\lower.7ex\hbox{E}\kern-.125emX}}

\usepackage{xcolor}
\usepackage{mathtools}
\usepackage{epsfig,makeidx,color}
\usepackage{subcaption}
\usepackage{amsmath,amssymb,bbm,enumitem,amsthm}
\usepackage{algorithm}
\usepackage{algcompatible}
\usepackage{cite,graphicx,lipsum}
\usepackage{makecell, array, tabularx, multicol, multirow, booktabs, boldline}

\usepackage[switch,pagewise]{lineno}
\usepackage{hyperref}
\usepackage[cjk]{kotex}
\hypersetup{
        colorlinks = true,
        citecolor=red,
}

\newcommand{\bX}{\mathbf{B}(X)}

\def\rT{{\rm T}}

\def\uR{{\mathbb R}}
\def\uC{{\mathbb C}}
\def\uE{{\mathbb E}}

\def\be{ \begin{equation} }
\def\ee{ \end{equation} }
\def\bea{ \begin{eqnarray} }
\def\eea{ \end{eqnarray} }

\def\bX{{\bf x}}

\def\bc{{\bf c}}

\def\bu{{\bf u}}

\def\bn{{\bf n}}

\def\bz{{\bf z}}

\def\be{{\bf e}}

\def\bA{{\bf A}}
\def\bB{{\bf B}}
\def\bC{{\bf C}}

\def\bI{{\bf I}}

\def\bX{{\bf X}}

\def\b0{{\bf 0}}

\def\cC{{\cal C}}

\def\cN{{\cal N}}

\ifCLASSOPTIONonecolumn
\else

\fi

\usepackage{geometry}
 \newcommand{\linebreakand}{%
  \end{@IEEEauthorhalign}
  \hfill\mbox{}\par
  \mbox{}\hfill\begin{@IEEEauthorhalign}
}
\makeatother

\begin{document}
\setlength{\columnsep}{0.2in}
\title{Hybrid Semantic-Complementary Transmission for High-Fidelity Image Reconstruction}

\author{Hyelin Nam$^1$, Jihong Park$^2$, Jinho Choi$^3$ and Seong-Lyun Kim$^1$ \\
$^1$School of Electrical and Electronic Engineering, Yonsei University, Seoul, South Korea,\\
$^2$Information Systems Technology and Design Pillar, Singapore University of Technology and Design, Singapore,\\
$^3$School of Electrical and Mechanical Engineering, The University of Adelaide, Adelaide, Australia,\\
Email: $^1$\{hlnam, slkim\}@ramo.yonsei.ac.kr, $^2$jihong\_park@sutd.edu.sg, $^3$jinho.choi@adelaide.edu.au}

\maketitle

\begin{abstract}
Recent advances in semantic communication (SC) have introduced neural network (NN)-based transceivers that convey semantic representation (SR) of signals such as images. However, these NNs are trained over diverse image distributions and thus often fail to reconstruct fine-grained image-specific details. To overcome this limited reconstruction fidelity, we propose an extended SC framework, hybrid semantic communication (HSC), which supplements SR with complementary representation (CR) capturing residual image-specific information. The CR is constructed at the transmitter, and is combined with the actual SC outcome at the receiver to yield a high-fidelity recomposed image. While the transmission load of SR is fixed due to its NN-based structure, the load of CR can be flexibly adjusted to achieve a desirable fidelity. This controllability directly influences the final reconstruction error, for which we derive a closed-form expression and the corresponding optimal CR. Simulation results demonstrate that HSC substantially reduces MSE compared to the baseline SC without CR transmission across various channels and NN architectures.
\end{abstract}


\section{Introduction}

Semantic communication (SC) has recently emerged as a novel paradigm in image transmission, shifting the focus from accurately conveying image pixels to effectively delivering their meanings or semantics \cite{10387520, 10559407, choi2024semantic, du2024objectattributerelationrepresentationbasedvideo}. Recent advancements in deep neural networks (NNs) have facilitated NN-based SC, in which the encoder transmits semantic representations (SRs) extracted from source images, while the decoder reconstructs images that are perceptually similar to the originals \cite{10446638, peng2024robustsemanticcommunicationimage}.

While SC offers remarkable advantages in bandwidth efficiency and robustness under poor channel conditions \cite{wang2024semanticcommunicationsexplicitsemantic, hu2025taskagnosticsemanticcommunicationmultimodal}, it remains difficult to flexibly control and enhance image reconstruction fidelity. Unlike classical communication, which can improve fidelity by increasing payload sizes, transmission power, or bandwidth allocation, SC lacks such flexibility due to its core objective of achieving perceptual similarity rather than exact pixel-level accuracy. Consequently, SRs often fail to capture fine-grained, sample-specific details, such as precise textures, colors, and spatial arrangements, thereby constraining reconstruction fidelity \cite{10960413, 10628028, cicchetti2024language}. Simply increasing SR sizes does not yield a proportional improvement in reconstruction fidelity. While leveraging multiple data modalities can partially improve fidelity \cite{cicchetti2024language}, they still fall short of increasing reconstruction fidelity up to the highest level, i.e., zero mean squared error (MSE) in image pixels.

To address this limitation, we propose hybrid semantic communication (HSC), which integrates SC with classical communication by jointly transmitting complementary representations (CRs) alongside the SRs. In HSC, the transmitter predicts the error after decoding SR alone and constructs the CR to compensate it. At the receiver, the received CR and decoded SR are projected onto range and null spaces, respectively, and are combined to recompose the final image reconstruction using the range-null space decomposition property of the original image \cite{grassucci2024rethinkingmultiusersemanticcommunications, ijcai2023p69, wang2022zeroshotimagerestorationusing}.

Assuming an error-free channel for theoretical analysis, we derive the optimal projection matrix for HSC that minimizes the MSE between the reconstructed and original image, leading to a closed-form solution for the CR. Importantly, this CR compensates for distortion caused by missing information from source images, rather than channel perturbations. Therefore, the optimal CR remains useful under non-ideal channels, provided that channel perturbations can be separately corrected. For SRs, this channel denoising can be trivially achieved via few-shot fine-tuning of the SC encoder and decoder for a given channel. To correct channel perturbations for CRs, we further introduce a neural-augmented CR (NCR) fine-tuning method, where a shallow autoencoder (AE) NN is employed for denoising the received CR.


Numerical simulations on the MNIST and Flickr8k datasets demonstrate that HSC consistently outperforms SC across various NN encoder-decoder architectures, including variational AE (VAE) and vector-quantized VAE (VQ-VAE), achieving up to $0.91$ times reduction in MSE while maintaining the same total payload size as SC in an error-free channel. Under a Rayleigh fading channel, HSC with NCR fine-tuning reduces 74\% MSE of its fading channel counterpart, with the same total payload size.

\begin{figure*}[t]
    \centering
    \includegraphics[clip,width=.8\textwidth]{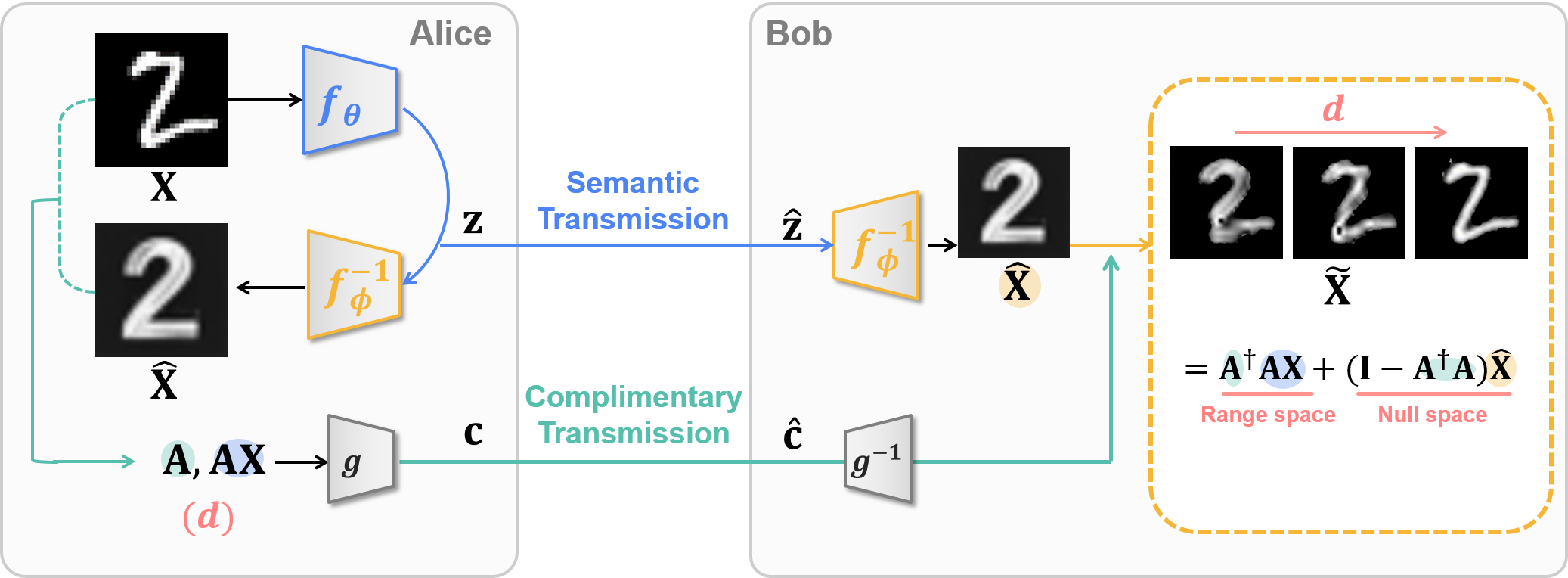}
    \caption{Illustration of Hybrid Semantic Communication with transmitting complementary representation.}
    \label{fig: instan_fig}
\end{figure*}

\section{VAE-based Hybrid Semantic Communication} \label{section: SM}
Our system architecture integrates two distinct transmission schemes: SR and CR transmissions, implemented through NN-based SC and conventional digital communication systems, respectively. In SC, a NN-structured transceiver is used to transmit the SR of the original image. A widely adopted approach is deep joint source and channel coding (DeepJSCC) \cite{8683463}, in which the encoder and the decoder are jointly trained in an end-to-end manner to perform multiple tasks, including source coding, channel coding, and SR pre/post-processing. This enables direct mapping of the original image to SR in a form on channel input signal while maintaining robustness under varying channel conditions. In particular, this work adopts a variational autoencoder (VAE)-based SC framework \cite{Kingma14, saidutta2021joint, 10683686, 10663301}, where the encoder transmits SR constrained by a certain distribution prior. The decoder reconstructs the image by sampling from the learned distribution.

The transmitter is equipped with $\theta$-parameterized semantic encoder, based on a VAE, mapping an original image $\bX \in \uR^{L\times L}$ into the SR $\bz\in \uC^k$, where $k<L^2$. This consists of three steps to produce channel input symbols to be directly transmitted: First, the encoder extracts two parallel NN output, $\bz_\mu, \bz_\sigma \in \uC^k$, each representing mean and variance features of the original image. Then they are reparameterized to produce a resampled vector following a Gaussian prior, $\bar \bz=\bz_\mu+\varepsilon\cdot \bz_\sigma$, where $\varepsilon\sim \cC\cN(0,1)$ is a standard complex Gaussian random variable. Finally, to meet the power constraint $P$, the sampled vector is normalized as $\bz=\sqrt{kP}\frac{\bar \bz}{\|\bar\bz\|_2}$, to ensure $\frac{1}{k}\uE\|\bz\|_2^2 \leq P$. The entire encoding process that produces SR in the form of $k$ channel input symbols is expressed as:
\begin{align}
    \bz= f_\theta (\bX), \quad \bz\in\uC^k.
\end{align}

Subsequently, Alice transmits the CR, denoted as $\bC \in \uR^r$, where $r$ and the method of generating $\bC$ are detailed in the following sections. The CR is transmitted using a conventional three-stage digital communication pipeline: source coding, channel coding, and modulation. These are unlike NN-based DeepJSCC transceivers for semantic transmission, which are trained to adapt to diverse channels. First, a source coding compresses $\bC$ into a bitstream using an image compression method such as JPEG or JPEG2000. This bitstream is then passed through a channel coding block to produce error-correcting codes (e.g., LDPC or Turbo codes). Finally, the encoded bitstream is modulated into complex-valued channel input symbols $\bc$, conforming to a selected modulation scheme. The entire process is represented by a non-trainable function $g: \uR^r \rightarrow \uC^{r\cdot R}$, where $R < 1$, expressed as:
\begin{align}
    \bc= g (\bC), \quad \bc\in\uC^{r\cdot R}.
\end{align}

The SR and CR are transmitted over two independent slow-fading channels, each subject to fading and additive white Gaussian noise (AWGN), with identical average channel gain and noise power. The received SR and CR at Bob are given by:
\begin{align}
    \hat \bz &= h_z \sqrt{P}\bz + \bn_z, \quad \hat\bz \in \uC^k, \\
    \hat \bc &= h_c \sqrt{P}\bc + \bn_c, \quad \hat\bc \in \uC^{r\cdot R}.
\end{align}
where $h_z, h_c \sim \cC\cN(0, \mu)$ is a modeled channel fading coefficient following a complex Gaussian distribution with variance $\mu$ representing the average channel gain. The additive noise vector $\bn_z\sim \cC\cN(0,\sigma^2\bI_z)$ and $\bn_c\sim \cC\cN(0,\sigma^2\bI_c)$ also follows random distribution with the noise power $\sigma^2$, determined by averaged signal-to-noise ratio (SNR) level, expressed as:
\begin{align} \label{eq: snr}
    \mathrm{SNR} = 10\log_{10} \frac{|h_z|^2 P}{\sigma^2} \,\, \mathrm{[dB]}.
\end{align}

Once receiving $\hat \bz$, Bob reconstructs an image from the received SR through the semantic decoder parameterized by $\phi$. Leveraging the decoder’s capability to generate realistic images from the assumed prior distribution, the generated image is perceptually similar with the original image, yet far from expressing the fine detail of the original image. The decoding process is represented as:
\begin{align}
    \hat \bX=f^{-1}_\phi (\hat \bz), \quad \hat \bX \in \uR^{L\times L}.
\end{align}

The received CR is processed into the inverse sequence of source and channel encoding. The received symbols are demodulated into a real-valued bitstream, then processed by channel decoding to correct channel errors. With source decoding scheme, it decompresses and reconstructs the CR $\hat \bC$, as the overall decoding function is expressed as:
\begin{align}
    \hat \bC = g^{-1}(\hat \bc),s \quad \hat \bC \in \uR^r.
\end{align}

Prior to downstream communication, semantic encoder and decoder—collectively referred to as the semantic model—are jointly trained to extract and interpret the SR for reconstructing a semantically similar image. This training is conducted over an idealized, error-free channel where the noise variance $\sigma^2 \rightarrow 0$ and the channel coefficient $h_z \rightarrow 1$, such that $\bz=\hat \bz$. Here, a training dataset consisting of $M$ pairs of original images, their corresponding SRs and generated images are utilized. The training objective is formulated based on the variational lower bound \cite{sutter2021generalizedmultimodalelbo, kingma2022autoencodingvariationalbayes}, which facilitates learning SR aligned with a Gaussian prior.

\section{High-Fidelity Image Reconstruction with Complementary Transmission}
The proposed HSC architecture extends the SC framework by introducing an additional transmission of CR, extracted from the original image. At the receiver, the CR is used to refine the generated image, resulting in a high-fidelity recomposed image, $\tilde \bX\in \uR^{L\times L}$, with reduced pixel-wise error. The additional transmitting load is determined by a flexibly controllable parameter $d$. Accordingly, the reconstruction error diminishes, as detailed in the following sections, showing that the transceiver can balance communication cost and reconstruction fidelity. In contrast, a naive alternative—directly transmitting the pixel-wise difference, $\bX-\hat \bX$, incurs a high transmission overhead, potentially approaching that of transmitting the original image. 

\subsection{Recomposing Image using Complementary subspaces}
Upon receiving the SR, Bob first generates an image, $\hat \bX$. By additionally utilizing the received CR, Bob enhances this image into a high-fidelity recomposed image, $\tilde \bX$. In particular, the recomposed image is obtained by integrating the generated image from the SR with the CR transmitted via the hybrid scheme from Alice, using complementary subspaces. We define a full-rank projection matrix, $\bA \in \uR^{d \times L}$, with its Moore–Penrose pseudoinverse $\bA^\dagger \in \uR^{L \times d}$, satisfying $\bA\bA^\dagger\bA = \bA$. This enables forming the range space, $\bA^\dagger \bA$, and its orthogonal complement, the null space, $\bI - \bA^\dagger \bA$, where $\bI$ is the identity matrix. Assuming Alice has a projection matrix $\bA$, she transmits the CR composed of $\bA$ and the projected image $\bA\bX$, where the dimension of $\bA$—proportional to the parameter $d$—determines both the transmission payload of the CR and, as will be explained later, the fidelity of the recomposed image. Bob yields a high-fidelity recomposed image, $\tilde \bX\in \uR^{L\times L}$, by combining the SR projected on the range space, and the CR projected on the null space, as:
\begin{align} \label{eq: RR}
    \tilde \bX & := \bA^\dagger\bA\bX + (\bI-\bA^\dagger\bA) \hat \bX.
\end{align}
With the same projection matrix, the following range-null decomposition holds with any original image $\bX$:
\begin{align} \label{eq: range}
    \bX & = \bA^\dagger\bA\bX + (\bI-\bA^\dagger\bA) \bX.
\end{align}

The objective of our method is to minimize the reconstruction error, measured by the Mean Squared Error (MSE), i.e.,
\begin{align} \label{eq: min}
    \min_\bA    ||\bX - \tilde\bX||^2.
\end{align}
To achieve a minimal reconstruction error, we aim to find the optimal $\bA$. Subtract \eqref{eq: RR} from \eqref{eq: range} to derive the objective function, \eqref{eq: min}:
\begin{align}
||\bX - \tilde\bX||^2 
&= \left\lVert \bA^\dagger\bA(\bX- \bX)  + (\bI - \bA^\dagger \bA)(\bX - \hat\bX) \label{eq: mse subtract} \right\rVert^2 \\
\label{eq:A_mse}
&= \left\lVert  (\bI - \bA^\dagger \bA)(\bX - \hat\bX) \right\rVert^2 \\
\label{eq:idempo}
&= (\bX - \hat\bX)^\rT(\bI - \bA^\dagger \bA)(\bX - \hat\bX),
\end{align}
where last step follows from the idempotent property of the symmetric projection matrix, i.e., $(\bI - \bA^\dagger \bA)^2=\bI - \bA^\dagger \bA$.

Let  $\bB\in\uR^{L\times L}$ be the error matrix, defined as the squared difference between the original and generated images. Then, the error matrix can be expressed using its eigenvalue decomposition as:
\begin{align}
    \bB = (\bX -\hat\bX)(\bX -\hat\bX)^\rT=\sum_{l=1}^{L}\lambda_l \be_l \be_l^T,
\end{align}
where $\lambda_l$ is the $l$-th largest eigenvalue, and $\be_l$ is its associated eigenvector. The MSE derivation in \eqref{eq:idempo} is continued with trace operator, denoted as `$\rm Tr$' which computes the sum of the diagonal elements of a square matrix. Its linearity property enables derivation:
\begin{align}
\label{eq:scalar}
&(\bX - \hat\bX)^\rT(\bI - \bA^\dagger \bA)(\bX - \hat\bX)
= {\rm Tr}\left(  (\bI - \bA^\dagger \bA)\bB  \right)\\
\label{eq:MSE_B}
&= {\rm Tr} (\bB) - {\rm Tr}(\bA^\dagger\bA\bB)\\
&=\sum_{l=1}^{L}\lambda_l - \sum_{l=1}^{L}\lambda_l \,{\rm Tr}(\bA^\dagger \bA \be_l\be_l^T)\\
&=\sum_{l=1}^{L}\lambda_l - \sum_{l=1}^{L}\lambda_l \,\be_l^T(\bA^\dagger \bA)\be_l\\
&=\sum_{l=1}^{L}\lambda_l - \sum_{l=1}^{L}\lambda_l \left\lVert\bA \be_l\right\rVert_2^2. \label{eq: Ae2}
\end{align}

Considering that $\bA$ is a linear operation mapping $\uR^{L\times L} \rightarrow \uR^{d\times L}$, where $d<L$, we have $\left\lVert\bA \be_l\right\rVert_2^2 \leq \left\lVert \be_l\right\rVert_2^2 = 1$. Here, the equality holds if and only if $\be_l$ spans the range space of $\bA$. This leads to \eqref{eq: Ae2} being bounded below by $\sum_{l=1}^{L}\lambda_l - \sum_{l=1}^{d}\lambda_l$. When $d=L$, the substitution yields an MSE of zero, while for $d\in [1, L-1]$, the minimum achievable MSE is given by $\sum_{l=d+1}^{L}\lambda_l$. The case $d=0$ corresponds to the SC framework, which transmits only the SR. The minimum MSE is achieved by $\bA$ consisting of the eigenvectors associated with the $d$-largest eigenvalues:
\begin{align}
\bA=  [\be_1 \ \cdots \ \be_d]^\rT. \label{Eq:Prop1}
\end{align}

\subsection{Transmitting complimentary representation}
This section presents a practical implementation of HSC, focusing on the transceiver design under channel impairments. To enable Bob to reconstruct the image as in \eqref{eq: RR}, Alice transmits the CR as $\bC = [\bA | \bA\bX] \in \uR^{2d \times L}$, where $\bA$ follows \eqref{Eq:Prop1}, and `$|$' denotes horizontal concatenation of the complementary matrix $\bA$ and the projected image $\bA\bX$, thus satisfying $r = 2d \times L$. Since the optimal projection matrix $\bA$ in \eqref{Eq:Prop1} is computed from the error matrix involving both $\bX$ and $\hat \bX$, Alice shares the same decoding function $f^{-1}_\phi$ as Bob to synchronize the generation of $\hat \bX$, as illustrated in Figure \ref{fig: instan_fig}.

Unlike pre-trained NN-based transceivers used for SR transmission, the CR transmission offers flexibility in controlling the transmission load. Alice achieves this by adjusting the parameter $d$, subject to the constraint that the combined payload of the SR and CR remains smaller than that of directly transmitting the original image. The compression ratio $\eta$ quantifies this relative transmission load. While SC uses a fixed NN-based semantic encoder $f_\theta:\uR^{L\times L} \rightarrow \uC^k$, the non-SC component, including the CR, is transmitted using a conventional digital scheme with a reduced rate $R$, as described in Section \ref{section: SM}. The resulting compression ratio is given by:
\begin{align} \label{eq: eta}
    \eta=\frac{|\bz|+|\bC|}{|\bX|\cdot R}=\frac{|\bz|+|\bA|\cdot R+|\bA\bX|\cdot R}{|\bX|\cdot R}=\frac{k/R+2dL}{L^2}.
\end{align}
To ensure $\eta < 1$, the following constraint on $d$ must be satisfied as: $d<L/2 - k/2LR$.

Under fading and noisy channel conditions, we propose distinct denoising strategies for the transmission of SR and CR. The SR is transmitted through a NN-based transceiver, pre-trained under an ideal and noise-free channel condition. However, when operating over a fading channel characterized by $h_z > 0$ and $\bn_z > 0$, the received SR becomes corrupted, leading to degraded reconstruction quality and even semantic inconsistency in the generated image. To mitigate this degradation, we adopt a few-shot fine-tuning strategy that adapts the transceiver to the target channel conditions. This is done by updating the parameters of the semantic encoder and decoder using a set of training samples collected under the actual channel, characterized by a fading coefficient $\mu=1$ and noise power $\sigma^2$ corresponding to an SNR range of 0–5 dB. The fine-tuned models, $f_{\hat\theta}$ and $f^{-1}_{\hat\phi}$, are obtained by minimizing the reconstruction error of the generated image, as formulated below:
\begin{align}
\{    \hat\theta, \hat\phi\}=\textrm{argmin}_{\hat\theta, \hat\phi}\uE[\lVert\bX-\hat \bX\rVert^2].
\end{align}

In contrast to the SR, the CR is transmitted through a deterministic digital scheme and therefore cannot be adapted to a fading channel (i.e., $h_c > 0$ and $\bn_c > 0$) using the few-shot adaptation strategy described earlier. To mitigate the distortion in the CR, we introduce a neural-augmented CR (NCR) fine-tuning scheme that incorporates shallow autoencoder-based adapters, $\tilde f_{\psi_d}$ and $\tilde f^{-1}_{\omega_d}$, into the semantic transceiver. These adapters are connected to the outputs of the fine-tuned semantic encoder and decoder. Their role is to indirectly mitigate the impact of channel corruption on CR by enabling the recomposed image $\tilde\bX$ to closely approximate the original image $\bX$. As a result, the output SR—processed through both the encoder and its adapter—becomes more robust to corrupted CR inputs, thereby enhancing the fidelity of the recomposed image. The adapters are trained by the objective formulated as:
\begin{align}
   \{ \psi_d, \omega_d\}=\textrm{argmin}_{\psi_d, \omega_d}\lVert\bX-\tilde \bX\rVert^2,
\end{align}
with the fine-tuned semantic model parameters $\hat\theta$ and $\hat\phi$ kept fixed. Each adapter is optimized independently for a given CR dimension by $d$. Combined with the few-shot fine-tuning of the SR transceiver, this dual strategy ensures that the proposed HSC framework remains resilient under fading channel conditions.

\begin{figure}[t!]
    \centering
    \includegraphics[clip,width=.9\columnwidth]{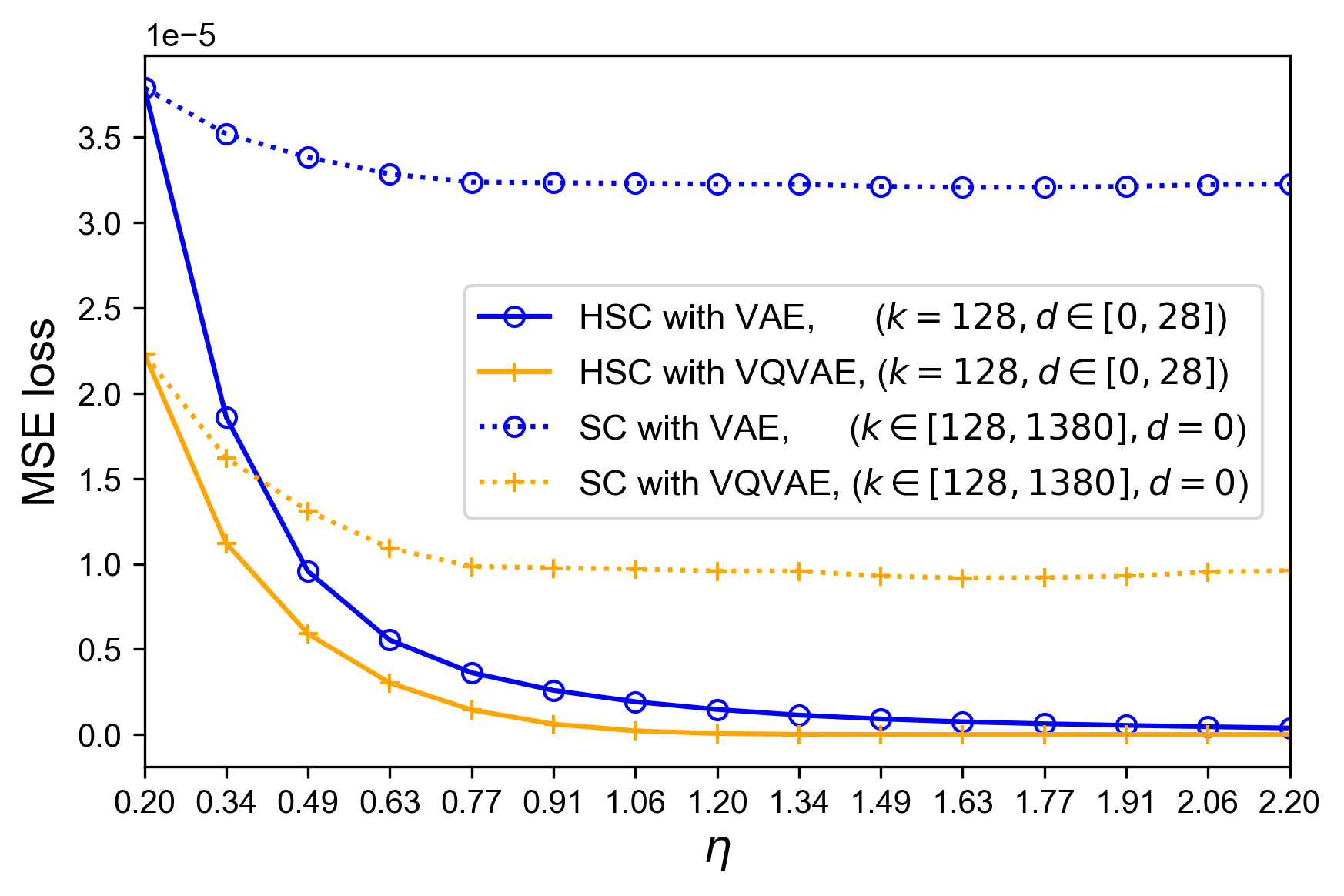}
    \caption{Comparison of transmitting SR (SC) and transmitting both SR and CR (HSC) with regard to compression ratio $\mu$. The experiments are done with VAE and VQVAE-based SC framework.}
    \label{fig: vae}
\end{figure}

\section{Numerical results}
\subsection{Experiment Settings}
For image recomposition based on linear matrix projections, we use 2-dimensional grayscale MNIST images \cite{deng2012mnist}. Each image has $28 \times 28$ ($L = 28$) pixels. A training set of $50,000$ images is used to train and fine-tune the semantic models, while a separate set of $10,000$ non-overlapping images is used for downstream transmission. The semantic encoder includes a flattening layer to transform the image dimension from $\mathbb{R}^{L \times L}$ to $\mathbb{R}^{L^2}$, followed by five fully-connected (FC) layers with output dimensions of $2048$, $1024$, $512$, and two parallel outputs of $k=128$. The semantic decoder consists of four FC layers, reversing the scale to output dimensions of $512$, $1024$, $2048$, and $784$, with the generated image reshaped back to $28 \times 28$. For complementary representations, $d$ is varied from $0$ to $L=28$ in integer steps. The CR is converted into channel input samples with a ratio of $R = 0.8$, compressed by $1/5$ using JPEG2000, followed by LDPC channel coding with a $0.5$ code rate and 16QAM modulation. In the fading channel, both SR and CR are affected by fading with an average channel gain $\mu = 1$ and additive noise with an average SNR ranging from $0$ to $5$ dB, increasing in steps of $0.5$ dB. The AE-based adapter for NCR fine-tuning includes two FC layers in both encoder and decoder. The adapter encoder maps the SR from size $128$ to $256$ and back to $128$, while the decoder consists of FC layers of sizes $1024$ and $784$, followed by reshaping to generate the image.

\begin{figure}[t]
    \centering
    \includegraphics[clip,width=.9\columnwidth]{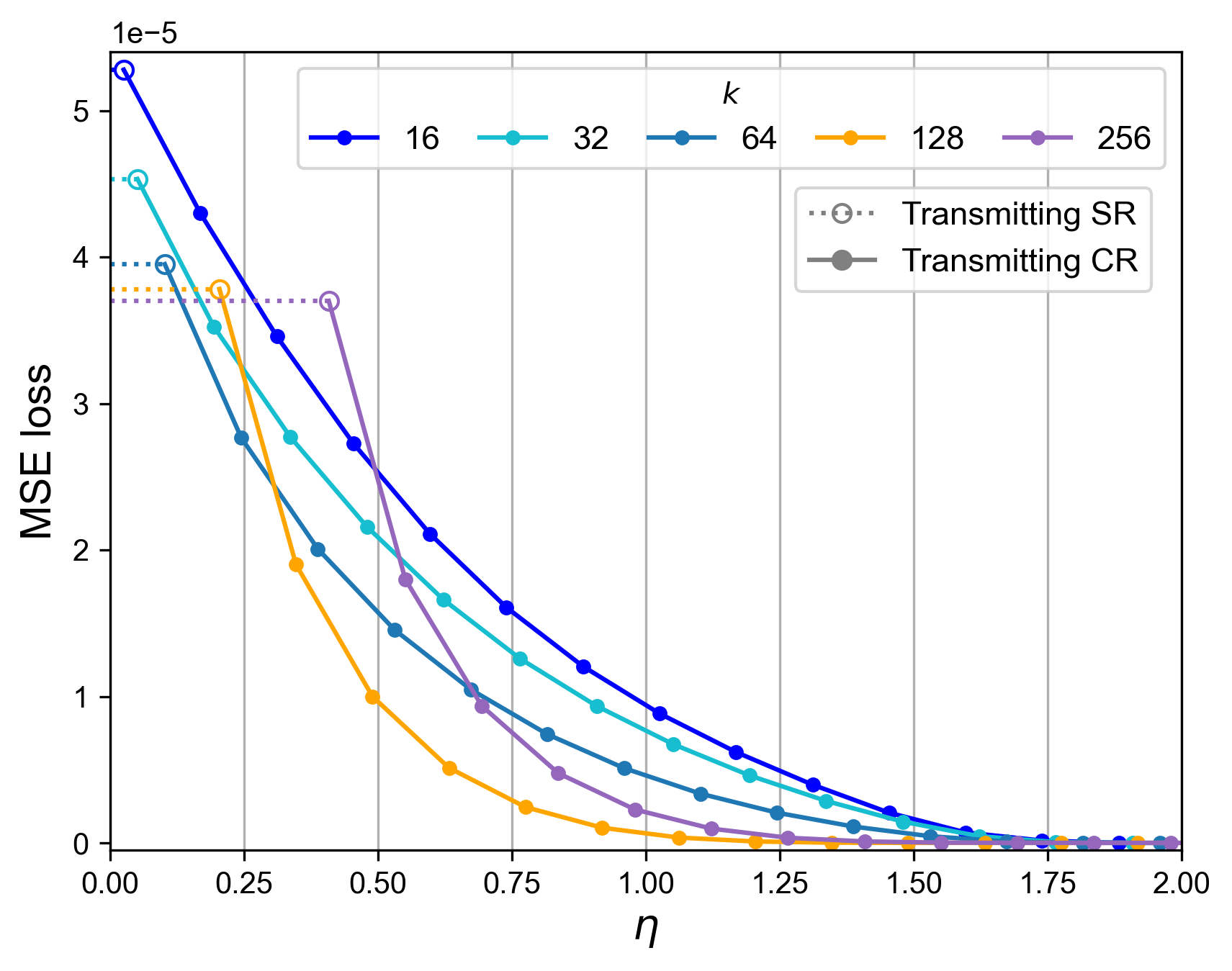}
    \caption{Reconstruction error of HSC under fixed total transmission load. Transmitting CR alongside a small SR yields lower MSE than transmitting SR alone with the same load.}
    \label{fig: CR}
\end{figure}

\subsection{Reconstruction error Reduction by transmitting CR compared to SR}
Figure \ref{fig: vae} compares the reconstruction error in an error-free channel when increasing the SR transmission within the SC framework and when increasing the CR transmission by the same amount in the HSC framework, under an equal transmission ratio $\mu$. In SC, $\mu$ is increased by enlarging $k$ while keeping $d=0$ in \eqref{eq: eta}, and the MSE is computed between the original and the generated images. In the HSC case, the transmitter fixes the SR at $k=128$ and increases the CR transmission by gradually raising $d$ in steps of 2. The reconstruction error is then measured using the recomposed images. Compared to increasing SR alone, adding CR more effectively reduces reconstruction error, achieving MSE = 0 when $d = L = 28$. In contrast, SR-only transmission causes the MSE to converge around $3.85 \times 10^{-5}$, even when its load exceeds that of the original image ($\eta > 1$). These findings indicate that HSC achieves higher communication efficiency than SC.

We also conducted an experiment with another SC framework using vector quantized variational autoencoder (VQVAE)-based transceivers \cite{10101778, choi2024semantic}. The NN architecture of the semantic encoder and decoder is follows that of the VAE-based SC, except for the parallel layer in the encoder. The VQVAE encoder outputs only $\bz_\mu \in \uC^k$, without reparameterization. Instead, $\bz_\mu$ is quantized by mapping it to the nearest vector $\bu_i \in \uC^k$ in a trainable codebook $\cC = \{\bu_1, \bu_2, \ldots, \bu_{256}\}$, resulting in $\bar \bz = \bu_i$, where $i := \textrm{argmin}_{\bu_i \in \cC} \lVert \bar \bz - \bu_i \rVert_2$. Then $\hat \bz$ is normalized to the SR $\bz$ to satisfy the transmitting power constraint. As shown in Figure \ref{fig: vae}, the VQVAE-based HSC also achieves a greater reduction in reconstruction error compared to transmitting the same amount of the SR alone.

Figure \ref{fig: CR} compares the reconstruction error of HSC for different SR transmission loads. The empty circles represent the reconstruction error between the original and generated images, while the filled circles indicate the reconstruction error of the recomposed images, which includes the transmission of CR within the SC framework. For transmission ratios larger than $\eta = 0.245$, transmitting CR results in a smaller MSE compared to increasing SR transmission.

\subsection{Denoising HSC in a fading channel}
\begin{figure}[t]
    \centering
    \includegraphics[clip,width=.9\columnwidth]{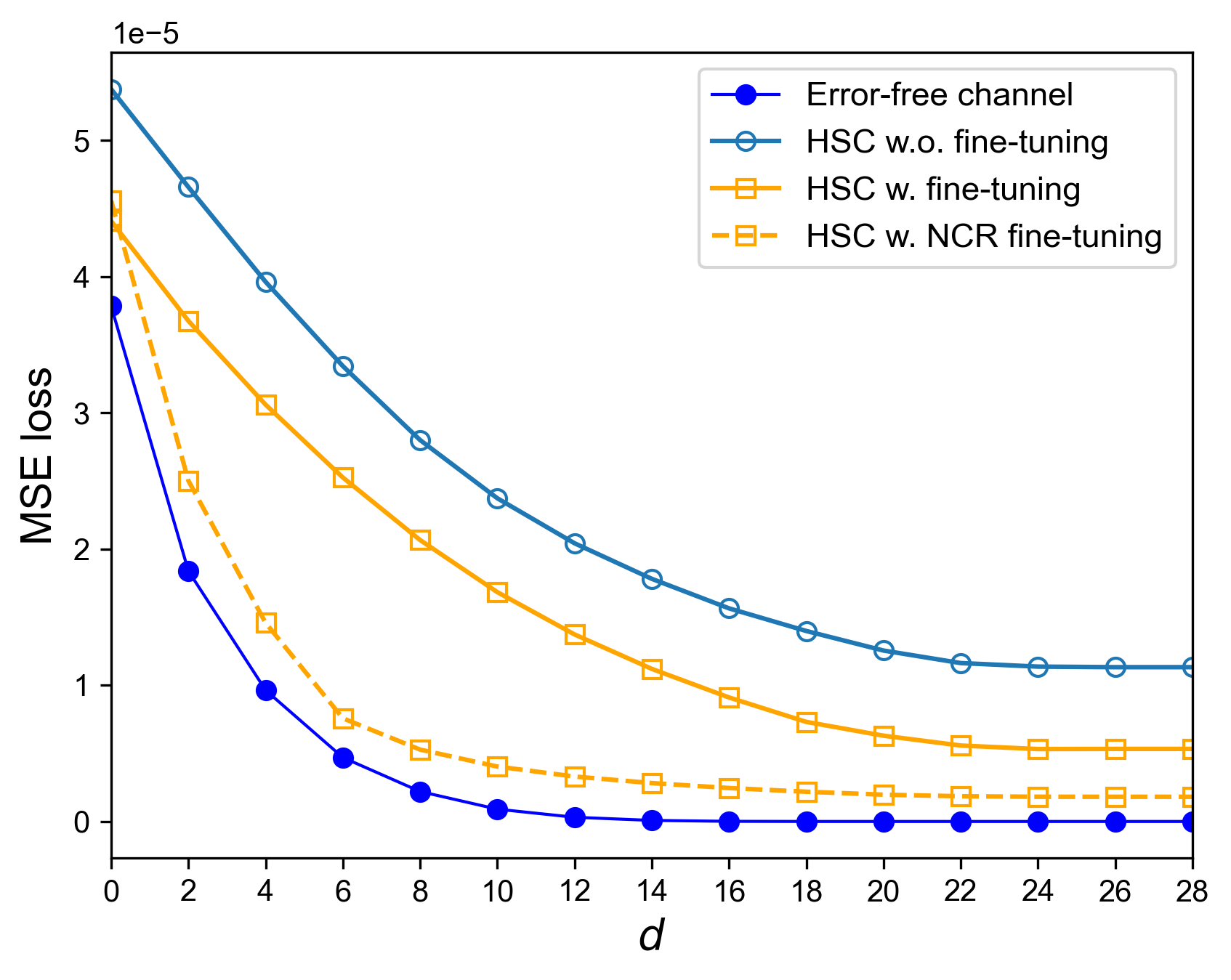}
    \caption{Reconstruction error with regard to $d$ in error-free channels and fading channels, demonstrating the error reduction using fine-tuning and NCR fine-tuning.}
    \label{fig: denois}
\end{figure}

\begin{figure}[t]
    \centering
    \includegraphics[clip,width=.9\columnwidth]{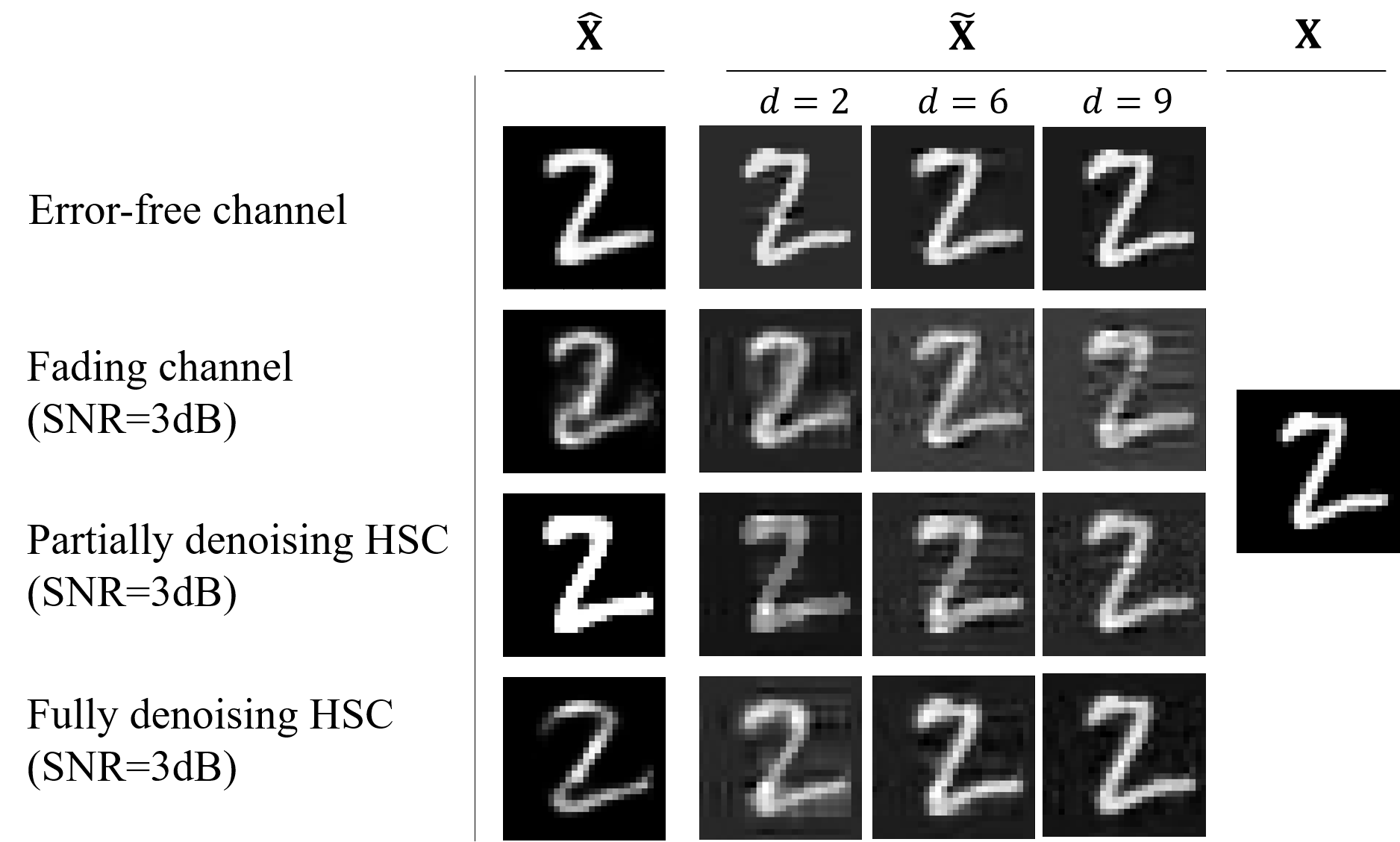}
    \caption{Example results for varying $d$. Generated images $\hat\bX$ and recomposed images $\tilde\bX$ show improved fidelity as $d$ increases.}
    \label{fig: ex_denois1}
\end{figure}

\begin{figure}[t]
    \centering
    \includegraphics[clip,width=.9\columnwidth]{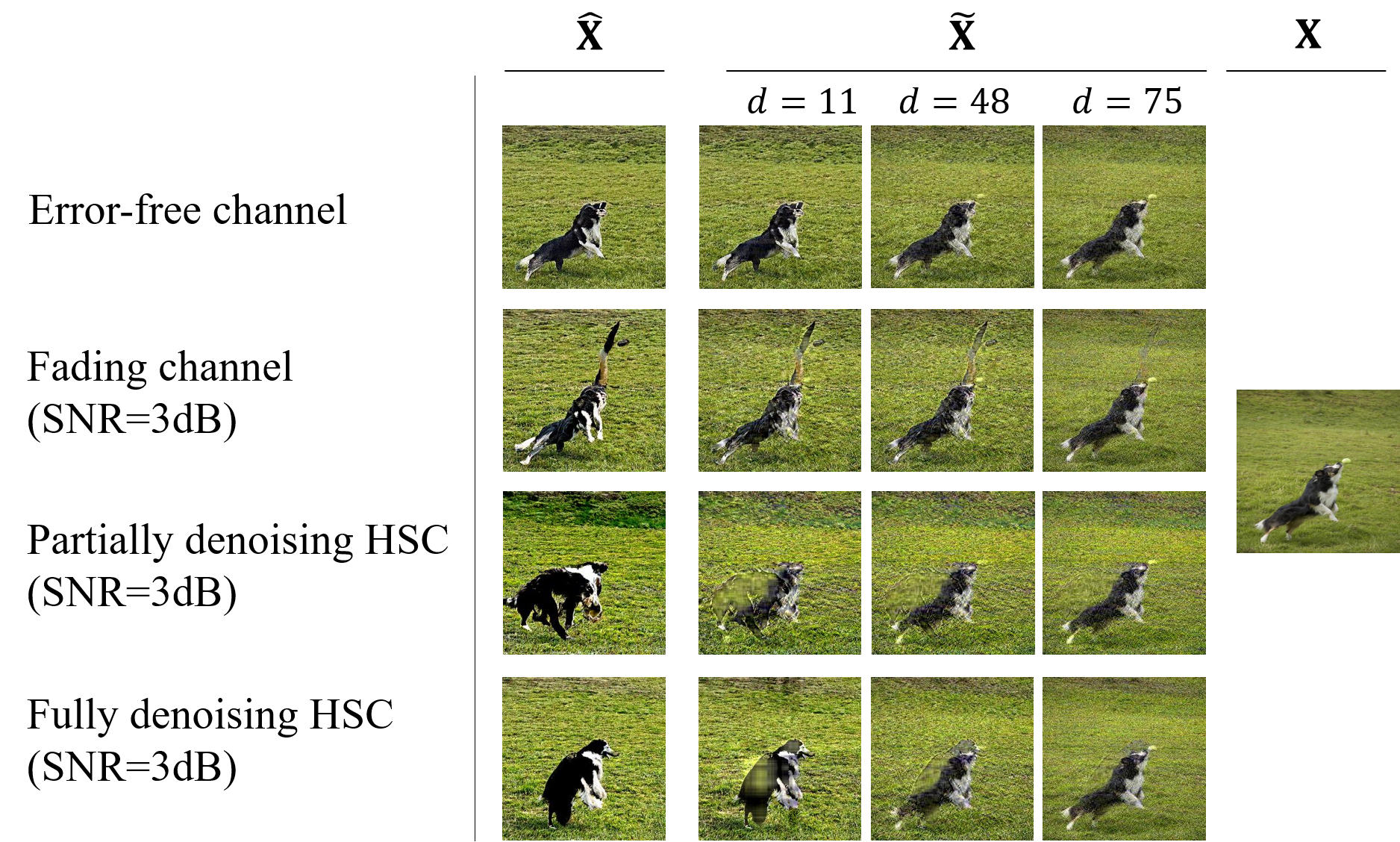}
    \caption{Example results for varying $d$, with 3-dimensional high resolutional image.}
    \label{fig: ex_denois2}
\end{figure}

In a fading and AWGN channel, semantic models trained for an error-free channel are inadequate, prompting the need for fine-tuning to produce denoising models, as Figure \ref{fig: denois} represents. Transmission of the SR and CR over a fading channel results in an increase in MSE of approximately $2\times 10^{-5}$ across all values of $d$, compared to the ideal channel case. Fine-tuning reduces the error by an average factor of $0.36$, while the use of adapters for NCR fine-tuning further lowers the MSE—even at small $d$—by an averaged $0.74$ ratio. Figures \ref{fig: ex_denois1} and \ref{fig: ex_denois2} present example results using an MNIST image and a high-resolution $256\times 256$ image from the Flickr8k dataset \cite{hodosh2013flickr}, respectively. The complementary projection matrix is obtained by averaging the error matrices computed from the three color channels of the original and generated images. The results illustrate that the generated images are affected by channel conditions and denoising strategies, while the recomposed images progressively recover finer details as $d$ increases.


\section{Conclusion}
This work addresses a key limitation of SC, namely that reconstruction fidelity saturates as the size of SR increases. To overcome this, we propose a novel framework of hybrid semantic communication (HSC), which enhances fidelity by additionally transmitting the CR alongside the SR, while balancing the trade-off between the transmission load and reconstruction error by explicitly controlling the CR's rank.  To further improve robustness under channel impairments, we introduce additional fine-tuning strategies which perform denoising SRs and/or CRs. Building upon this, it could be an interesting to study CR compression technqiues as well as incorporate advanced generative models.

\section*{Acknowledgment}
This work was supported in part by Institute of Information \& communications Technology Planning \& Evaluation (IITP) grant funded by the Korea government (MSIT) (No.2021-0-00347, 6G Post-MAC (POsitioning- \& Spectrum-aware intelligenT MAC for Computing \& Communication Convergence)) and in part by SUTD Kickstarter Initiative (SKI 2021\_06\_08).

\bibliographystyle{ieeetr}
\bibliography{Gobecom_ref}

\begin{thebibliography}{10}

\bibitem{10387520}
Z.~Qin, J.~Ying, D.~Yang, H.~Wang, and X.~Tao, ``Computing networks enabled semantic communications,'' {\em IEEE Network}, vol.~38, no.~2, pp.~122--131, 2024.

\bibitem{10559407}
J.~Choi, J.~Park, S.-W. Ko, J.~Choi, M.~Bennis, and S.-L. Kim, ``Semantics alignment via split learning for resilient multi-user semantic communication,'' {\em IEEE Transactions on Vehicular Technology}, vol.~73, no.~10, pp.~15815--15819, 2024.

\bibitem{choi2024semantic}
J.~Choi, J.~Park, E.~Grassucci, and D.~Comminiello, ``Semantic communication challenges: Understanding dos and avoiding don'ts,'' in {\em 2024 IEEE 99th Vehicular Technology Conference (VTC2024-Spring)}, pp.~1--5, IEEE, 2024.

\bibitem{du2024objectattributerelationrepresentationbasedvideo}
Q.~Du, Y.~Duan, Q.~Yang, X.~Tao, and M.~Debbah, ``Object-attribute-relation representation based video semantic communication,'' 2024.

\bibitem{10446638}
H.~Nam, J.~Park, J.~Choi, M.~Bennis, and S.-L. Kim, ``Language-oriented communication with semantic coding and knowledge distillation for text-to-image generation,'' in {\em ICASSP 2024 - 2024 IEEE International Conference on Acoustics, Speech and Signal Processing (ICASSP)}, pp.~13506--13510, 2024.

\bibitem{peng2024robustsemanticcommunicationimage}
X.~Peng, Z.~Qin, X.~Tao, J.~Lu, and K.~B. Letaief, ``A robust semantic communication system for image,'' 2024.

\bibitem{wang2024semanticcommunicationsexplicitsemantic}
F.~Wang, Y.~Zheng, W.~Xu, J.~Liang, and P.~Zhang, ``Semantic communications with explicit semantic bases: Model, architecture, and open problems,'' 2024.

\bibitem{hu2025taskagnosticsemanticcommunicationmultimodal}
J.~Hu, H.~Wu, W.~Zhang, F.~Wang, W.~Xu, H.~Gao, and D.~Gündüz, ``Task-agnostic semantic communication with multimodal foundation models,'' 2025.

\bibitem{10960413}
C.~Xu, M.~B. Mashhadi, Y.~Ma, R.~Tafazolli, and J.~Wang, ``Generative semantic communications with foundation models: Perception-error analysis and semantic-aware power allocation,'' {\em IEEE Journal on Selected Areas in Communications}, pp.~1--1, 2025.

\bibitem{10628028}
J.~Ren, Z.~Zhang, J.~Xu, G.~Chen, Y.~Sun, P.~Zhang, and S.~Cui, ``Knowledge base enabled semantic communication: A generative perspective,'' {\em IEEE Wireless Communications}, vol.~31, no.~4, pp.~14--22, 2024.

\bibitem{cicchetti2024language}
G.~Cicchetti, E.~Grassucci, J.~Park, J.~Choi, S.~Barbarossa, and D.~Comminiello, ``Language-oriented semantic latent representation for image transmission,'' in {\em 2024 IEEE 34th International Workshop on Machine Learning for Signal Processing (MLSP)}, pp.~1--6, IEEE, 2024.

\bibitem{grassucci2024rethinkingmultiusersemanticcommunications}
E.~Grassucci, J.~Choi, J.~Park, R.~F. Gramaccioni, G.~Cicchetti, and D.~Comminiello, ``Rethinking multi-user semantic communications with deep generative models,'' 2024.

\bibitem{ijcai2023p69}
X.~Cheng, N.~Zhang, J.~Yu, Y.~Wang, G.~Li, and J.~Zhang, ``Null-space diffusion sampling for zero-shot point cloud completion,'' in {\em Proceedings of the Thirty-Second International Joint Conference on Artificial Intelligence, {IJCAI-23}} (E.~Elkind, ed.), pp.~618--626, International Joint Conferences on Artificial Intelligence Organization, 8 2023.
\newblock Main Track.

\bibitem{wang2022zeroshotimagerestorationusing}
Y.~Wang, J.~Yu, and J.~Zhang, ``Zero-shot image restoration using denoising diffusion null-space model,'' 2022.

\bibitem{8683463}
E.~Bourtsoulatze, D.~B. Kurka, and D.~Gündüz, ``Deep joint source-channel coding for wireless image transmission,'' in {\em ICASSP 2019 - 2019 IEEE International Conference on Acoustics, Speech and Signal Processing (ICASSP)}, pp.~4774--4778, 2019.

\bibitem{Kingma14}
D.~P. Kingma and M.~Welling, ``{Auto-Encoding Variational Bayes},'' in {\em 2nd International Conference on Learning Representations, {ICLR} 2014, Banff, AB, Canada, April 14-16, 2014, Conference Track Proceedings}, 2014.

\bibitem{saidutta2021joint}
Y.~M. Saidutta, A.~Abdi, and F.~Fekri, ``Joint source-channel coding over additive noise analog channels using mixture of variational autoencoders,'' {\em IEEE Journal on Selected Areas in Communications}, vol.~39, no.~7, pp.~2000--2013, 2021.

\bibitem{10683686}
W.~Zhang, S.~Wu, S.~Meng, J.~He, and Q.~Zhang, ``Engineering a lightweight deep joint source-channel-coding-based semantic communication system,'' {\em IEEE Internet of Things Journal}, vol.~12, no.~1, pp.~458--471, 2025.

\bibitem{10663301}
F.~Wang, X.~Chen, and X.~Deng, ``Wireless adaptive image transmission over ofdm channels based on entropy model,'' {\em IEEE Wireless Communications Letters}, vol.~13, no.~10, pp.~2902--2906, 2024.

\bibitem{sutter2021generalizedmultimodalelbo}
T.~M. Sutter, I.~Daunhawer, and J.~E. Vogt, ``Generalized multimodal elbo,'' 2021.

\bibitem{kingma2022autoencodingvariationalbayes}
D.~P. Kingma and M.~Welling, ``Auto-encoding variational bayes,'' 2022.

\bibitem{deng2012mnist}
L.~Deng, ``The mnist database of handwritten digit images for machine learning research,'' {\em IEEE Signal Processing Magazine}, vol.~29, no.~6, pp.~141--142, 2012.

\bibitem{10101778}
Q.~Hu, G.~Zhang, Z.~Qin, Y.~Cai, G.~Yu, and G.~Y. Li, ``Robust semantic communications with masked vq-vae enabled codebook,'' {\em IEEE Transactions on Wireless Communications}, vol.~22, no.~12, pp.~8707--8722, 2023.

\bibitem{hodosh2013flickr}
M.~Hodosh, P.~Young, and J.~Hockenmaier, ``Framing image description as a ranking task: data, models and evaluation metrics,'' {\em J. Artif. Int. Res.}, vol.~47, p.~853–899, May 2013.

\end{thebibliography}

\end{document}